\documentclass[american,english,10pt,conference]{IEEEtran}
\usepackage[T1]{fontenc}
\usepackage[latin9]{inputenc}
\usepackage[letterpaper]{geometry}
\usepackage{verbatim}
\usepackage{float}
\usepackage{calc}
\usepackage{amsthm}
\usepackage{amsmath}
\usepackage{graphicx}
\usepackage{amssymb}

\makeatletter

\floatstyle{ruled}
\newfloat{algorithm}{tbp}{loa}
\floatname{algorithm}{Algorithm}

\theoremstyle{plain}
\newtheorem{thm}{Theorem}
\theoremstyle{definition}
\newtheorem{defn}[thm]{Definition}
\theoremstyle{plain}
\newtheorem{cor}[thm]{Corollary}
\theoremstyle{plain}
\newtheorem{lem}[thm]{Lemma}
\theoremstyle{remark}
\newtheorem{rem}[thm]{Remark}

 \usepackage{bbm}

\DeclareMathOperator*{\argmin}{arg\,min}

\date{}

\IEEEoverridecommandlockouts

\makeatother

\usepackage{babel}

\begin{document}

\title{An Iterative Joint Linear-Programming Decoding of LDPC Codes and
Finite-State Channels }

\author{\authorblockN{Byung-Hak Kim and Henry D. Pfister%
\thanks{This material is based upon work supported by the National Science
Foundation under Grant No. 0747470. Any opinions, findings, conclusions,
or recommendations expressed in this material are those of the authors
and do not necessarily reflect the views of the National Science Foundation.%
}} \authorblockA{Department of Electrical and Computer Engineering,
Texas A\&M University\\
 Email: \{bhkim,hpfister\}@tamu.edu} }
\maketitle
\begin{abstract}
In this paper, we \foreignlanguage{american}{introduce an efficient
iterative solver }for the joint linear-programming (LP) decoding of
low-density parity-check (LDPC) codes and finite-state channels (FSCs).
In particular, we extend the approach of iterative approximate LP
decoding, proposed by Vontobel and Koetter and explored by Burshtein,
to this problem. By taking advantage of the dual-domain structure
of the joint decoding LP, \foreignlanguage{american}{we obtain a convergent
iterative algorithm for joint LP decoding whose structure is similar
to BCJR-based turbo equalization (TE). The result is a joint iterative
decoder whose complexity is similar to TE but whose performance is
similar to joint LP decoding. The main advantage of this decoder is
that it appears to provide the predictability of joint LP decoding
and superior performance with the computational complexity of TE. }
\end{abstract}

\section{Introduction \label{sec:Intro} \vspace{0mm}}

Iterative decoding of error-correcting codes, while introduced by
Gallager in his 1960 Ph.D. thesis, was largely forgotten until the
1993 discovery of turbo codes by Berrou, et al. Since then, message-passing
iterative decoding has been a very popular decoding algorithm in research
and practice. In 1995, the turbo decoding of a finite-state channel
(FSC) and a convolutional code (instead of two convolutional codes)
was introduced by Douillard, et al as a \emph{turbo equalization}
(TE) \foreignlanguage{american}{which enabled the joint-decoding of
the channel and code by iterating between these two decoders} \cite{Douillard-ett95}.
Before this, one typically separated channel decoding from error-correcting
code decoding \cite{Gallager-60}\cite{Muller-it04}. This breakthrough
received immediate interest from the magnetic recording community,
and TE was applied to magnetic recording channels by a variety of
authors (e.g., \foreignlanguage{american}{\cite{Ryan-icc98,McPheters-asilo98,Oberg-aller98,Tuchler-com02}}).
TE \foreignlanguage{american}{was later combined with turbo codes}
and also extended to low-density parity-check (LDPC) codes (and called
\emph{joint iterative decoding}) by constructing one large graph representing
the constraints of both the channel and the code (e.g., \foreignlanguage{american}{\cite{Kurkoski-it02}}).

In \cite{Feldman-2003}\cite{Feldman-it05}, Feldman, et al. introduced
a linear-programming (LP) decoder for general binary linear codes
and considered it specifically for both LDPC and turbo codes. It is
based on solving an LP relaxation of an integer program which is equivalent
to maximum-likelihood (ML) decoding. For long codes and/or low SNR,
the performance of LP decoding appears to be slightly inferior to
belief-propagation decoding. But, unlike the iterative decoder, the
LP decoder either detects a failure or outputs a codeword which is
guaranteed to be the ML codeword. 

\begin{figure}[t]
\begin{centering}
\includegraphics[scale=0.65]{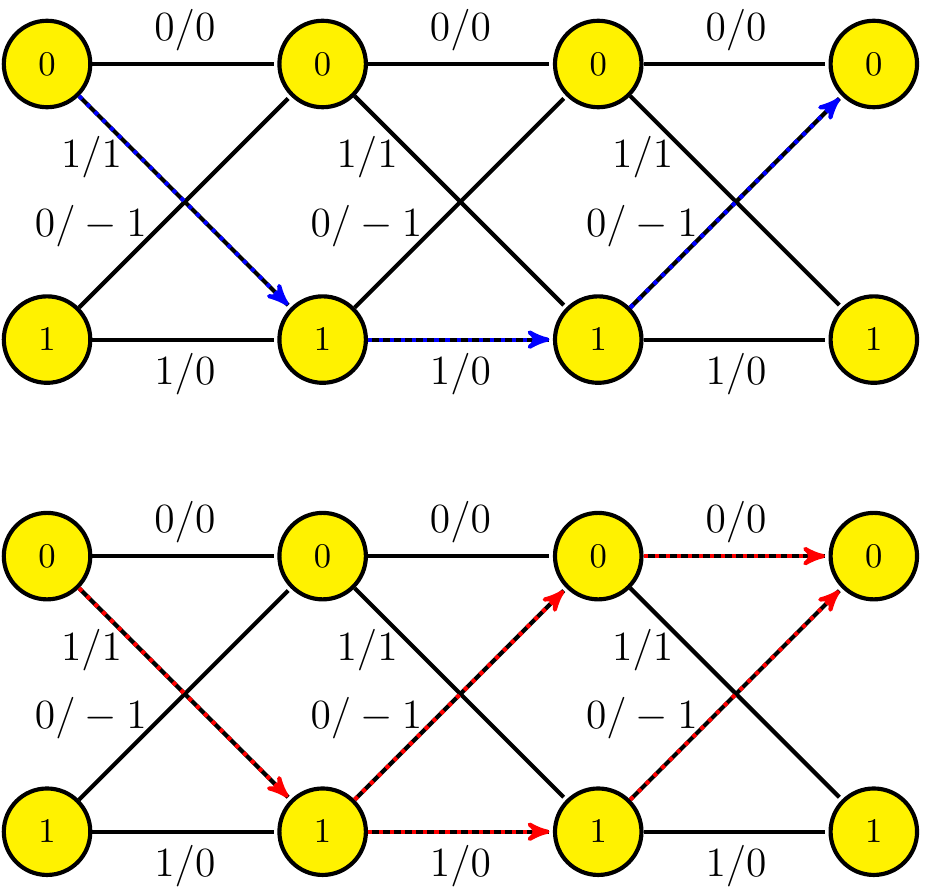}
\par\end{centering}

\caption{\label{fig:Example}Illustration of joint-LP decoder outputs for the
single parity-check code SPC(3,2) over DIC (starts in zero state).
By ordering the trellis edges appropriately, joint-LP decoder converges
to either a \emph{trellis-wise (ML) codeword} $(0\,1\,0\,0;0\,0\,0\,1;.0\,0\,1\,0)$
(top dashed blue path) or a \emph{joint-decoding trellis-wise pseudo-codeword}
$(0\,1\,0\,0;0\,0\,.5\,.5;.5\,0\,.5\,0)$ (bottom dashed red paths).
Using $\mathcal{Q}$ to project them into $\mathcal{P}(H)$, we obtain
the corresponding \emph{(symbol-wise) codeword} $(1,1,0)$ and \emph{joint-decoding
symbolwise pseudo-codeword} $(1,.5,0).$ }

\end{figure}

Recently, the LP formulation has been extended to the joint decoding
of a binary-input FSC and outer LDPC code \cite{Kim-isit10}\cite{Flanagan-arxiv09}.
In this case, the performance of LP decoding appears to outperform
belief-propagation decoding at moderate SNR. Moreover, all integer
solutions are indeed codewords and the joint decoder also has a certain
ML certificate property. This allows all decoder failures to be explained
by \emph{joint-decoding pseudo-codewords} \foreignlanguage{american}{(see
Fig. \ref{fig:Example})}. 

In the past, the primary value of LP decoding was as an analytical
tool that allowed one to better understand iterative decoding and
its modes of failure. This is because LP decoding based on standard
LP solvers is quite impractical and has a superlinear complexity in
the block length. This motivated several authors to propose low-complexity
algorithms for LP decoding of LDPC codes in the last five years \foreignlanguage{american}{(e.g.,
\cite{Wadayama-it10,Vontobel-turbo06,Vontobel-ita08,Taghavi-it08,Wadayama-isit09,Burshtein-it09,Punekar-aller10}).
Many of these have their roots in the iterative Gauss-Seidel approach
proposed by Vontobel and Koetter for approximate LP decoding \cite{Vontobel-turbo06}.
This approach was also studied further by Burshtein \cite{Burshtein-it09}. }

%
{}

In this paper, we extend this approach to the problem of low-complexity
joint LP decoding of LDPC codes and FSCs. We argue that by taking
advantage of the special structure in dual-domain of the joint LP
problem and replacing minima in the formulation with soft-minima,
we can obtain an efficient method that solves the joint LP. While
there are many ways to iteratively solve the joint LP, our main goal
was to derive one as the natural analogue of TE. This should lead
to an efficient method for joint LP decoding whose performance is
similar to joint LP and whose complexity similar to TE. \foreignlanguage{american}{Indeed,
the solution we provide is a fast, iterative, and provably convergent
form of TE and update rules are tightly connected to BCJR-based TE.
This demonstrates that an iterative joint LP solver with a similar
computational complexity as TE is feasible (see Remark \ref{rem:TEconnection}). }

%
{}%
{} 

The paper is structured as follows. After briefly reviewing joint
LP decoding in Sec. \ref{sec:Background}, Sec. \ref{sec:Joint-LPD}
is devoted to develop the iterative solver for the joint LP decoder\foreignlanguage{american}{,
i.e., iterative joint LP decoder and its proof of convergence}. Finally,
we provide, in Sec. \ref{sec:Simulations}, the decoder performance
results and conclude in Sec. \ref{sec:Concl}.\vspace{0mm} Due to
space limitations we omit many of the proofs, but they can be found
in \cite{Kim-stspsub11}.

\section{Background: Joint LP Decoder \label{sec:Background}}

\selectlanguage{american}%

\subsection{Notation}

Throughout the paper we borrow notation from \cite{Feldman-it05}.
Let $\mathcal{I}=\left\{ 1,\,\ldots,\, N\right\} $ and $\mathcal{J}=\left\{ 1,\,\ldots,\, M\right\} $
be sets of indices for the variable and parity-check nodes of a binary
linear code. A variable node $i\in\mathcal{I}$ is connected to the
set $\mathcal{N}(i)$ of neighboring parity-check nodes. Abusing notation,
we also let $\mathcal{N}(j)$ be the neighboring variable nodes of
a parity-check node $j\in\mathcal{J}$ when it is clear from the context.
For the trellis associated with a FSC, we let $E=\left\{ 1,\,\ldots,\, O\right\} $
index the set of trellis edges associated with one trellis section.
For each edge%
\footnote{In this paper, $e$ is used to denote a trellis edge while $\texttt{e}$
denotes the universal constant that satisfies $\ln\texttt{e}=1$.%
}, $e\in E^{N}$, in the length-$N$ trellis, the functions $t(e)\rightarrow\{1,\ldots,N\}$,~$s_{e}\rightarrow\mathcal{S}$,
$s'(e)\rightarrow\mathcal{S}$,~$x(e)\rightarrow\{0,1\}$,~and $a_{e}\rightarrow\mathcal{A}$
map this edge to its respective time index, initial state, final state,
input bit, and noiseless output symbol. Finally, the set of edges
in the trellis section associated with time $i$ is defined to be
$\mathcal{T}_{i}=\left\{ e\in E^{N}\,|\, t(e)=i\right\} $. %
\begin{figure}[t]
\begin{centering}
\includegraphics[scale=0.65]{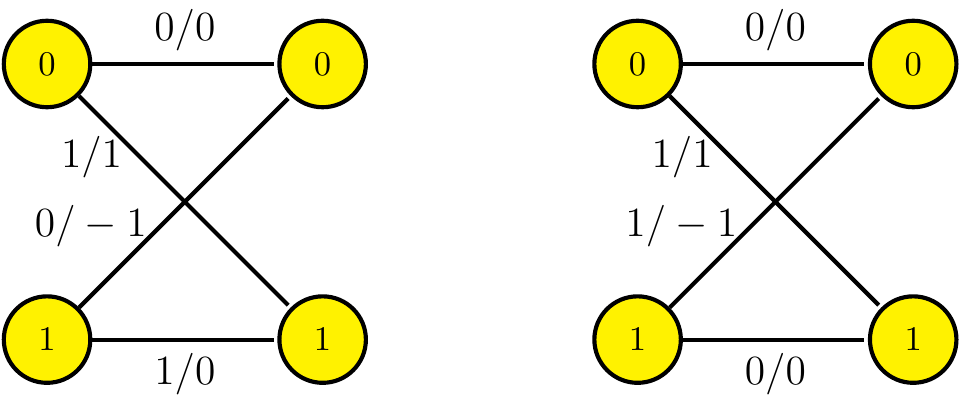}
\par\end{centering}

\caption{\label{fig:dic}The \emph{dicode channel} (DIC) is a binary-input
FSC with a linear response of $G(z)=1-z^{-1}$ and additive Gaussian
noise. If the input bits are differentially encoded prior to transmission,
then the resulting channel is called the \emph{precoded dicode channel}
(pDIC). The state diagrams of these two channels are shown: noiseless
dicode channel without (left) and with precoding (right). The edges
are labeled by the input/output pair.}

\end{figure}

\selectlanguage{english}%

\subsection{Joint LP Decoder}

\selectlanguage{american}%
Now, we describe \emph{the joint LP decoder} in terms of the trellis
of the FSC and the checks in the binary linear code%
\footnote{Extensions of this joint LP decoder to non-binary linear codes is
straightforward based on \cite{Flanagan-it09}.%
}. Let $N$ be the length of the code and $\mathbf{y}=(y_{1},y_{2},\ldots,y_{N})$
be the received sequence. The trellis consists of $(N+1)|\mathcal{S}|$
vertices (i.e., one for each state and time) and a set of at most
$2N|\mathcal{S}|^{2}$ edges (i.e., one edge for each input-labeled
state transition and time). The LP formulation requires one indicator
variable for each edge $e\in\mathcal{T}_{i}$, and we denote that
variable by $g_{i,e}$. So, $g_{i,e}$ is equal to 1 if the candidate
path goes through the edge $e$ in $\mathcal{T}_{i}$. Likewise, the
LP decoder requires one cost variable for each edge and we associate
the branch metric $b_{i,e}$ with the edge $e$ given by\[
b_{i,e}\negthinspace\triangleq\negthinspace\begin{cases}
-\ln P\negthinspace\left(y_{t(e)},s'(e)|x(e),s(e)\right)\!\!\! & \mbox{if}\, t(e)\!>\!1\\
-\ln\left[P\left(y_{t(e)},s'(e)|x(e),s(e)\right)\! P_{0}\left(s(e)\right)\right]\!\!\! & \mbox{if}\, t(e)\!=\!1.\end{cases}\]

\begin{defn}
The \emph{trellis polytope} $\mathcal{T}$ enforces the flow conservation
constraints for channel decoder. The flow constraint for state $k$
at time $i$ is given by\[
\mathcal{F}_{i,k}\triangleq\left\{ \mathbf{g}\in[0,1]^{N\times O}\left|\,\sum_{\substack{e:s'(e)=k}
}g_{i,e}=\sum_{\substack{e:s(e)=k}
}g_{i+1,e}\right.\right\} .\]
Using this, the \emph{trellis polytope} $\mathcal{T}$ is given by\[
\mathcal{T}\triangleq\left\{ \mathbf{g}\in\bigcap_{i=1}^{N-1}\bigcap_{k\in\mathcal{S}}\mathcal{F}_{i,k}\left|\,\sum_{\substack{e\in\mathcal{T}_{p}}
}g_{p,e}=1,\,\mbox{for any}\, p\in\mathcal{I}\right.\right\} .\]

\end{defn}
\selectlanguage{english}%

\selectlanguage{american}%
\begin{defn}
\label{def:Projection}Let \emph{$\mathcal{Q}$ }be the projection
of $\mathbf{g}$ onto the input vector $\mathbf{f}=\left(f_{1},\ldots,f_{N}\right)\in[0,1]^{N}$
defined by $\mathbf{f}=\mathcal{Q}\mathbf{g}$ with \[
f_{i}=\sum_{\substack{e\in\mathcal{T}_{i}:\, x(e)=1}
}g_{i,e}.\]

\end{defn}
\selectlanguage{english}%

\selectlanguage{american}%
Let $\mathcal{C}\subseteq\left\{ 0,1\right\} ^{N}$ be the length-$N$
binary linear code defined by a parity-check matrix and $\mathbf{c}=(c_{1},\ldots,c_{N})$
be a codeword. Let $\mathcal{L}$ be the set whose elements are the
sets of indices involved in each parity check, or \[
\mathcal{L}=\left\{ \mathcal{N}(j)\subseteq\{1,\ldots,N\}|\, j\in\mathcal{J}\right\} .\]
Then, we can define the set of codewords to be\[
\mathcal{C}=\left\{ \mathbf{c}\in\left\{ 0,1\right\} ^{N}\,\bigg|\,\sum_{i\in L}c_{i}\equiv0\,\bmod2,\,\forall\, L\in\mathcal{L}\right\} .\]
The \emph{codeword polytope} is the convex hull of $\mathcal{C}$.
This polytope can be quite complicated to describe though, so instead
one constructs a simpler polytope using local constraints. Each parity-check
$L\in\mathcal{L}$ defines a local constraint equivalent to the extreme
points of a polytope in $\left[0,1\right]^{N}$.
\begin{defn}
\label{def:LCP} The \emph{local codeword polytope} $\mbox{LCP(\ensuremath{L})}$
associated with a parity check is the convex hull of the bit sequences
that satisfy the check. It is given explicitly by\[
\mbox{LCP}(L)\triangleq\!\!\bigcap_{\substack{S\subseteq L\\
\left|S\right|\text{odd}}
}\left\{ \mathbf{c}\in[0,\,1]^{N}\,\bigg|\sum_{i\in S}c_{i}-\!\!\!\sum_{i\in L-S}\!\! c_{i}\leq\left|S\right|\!-\!1\right\} .\]

\end{defn}
\selectlanguage{english}%

\selectlanguage{american}%
\begin{defn}
The \emph{relaxed polytope} $\mathcal{P}(H)$ is the intersection
of the LCPs over all checks and \begin{align*}
\mathcal{P}(H) & \triangleq\bigcap_{L\in\mathcal{L}}\mbox{LCP}(L).\end{align*}

\end{defn}
\selectlanguage{english}%

\selectlanguage{american}%
\begin{defn}
\label{def:TPOLY}The \emph{trellis-wise relaxed polytope} $\mathcal{P}_{\mathcal{T}}(H)$
for $\mathcal{P}(H)$ is given by 

\[
\mathcal{P}_{\mathcal{T}}(H)\triangleq\left\{ \mathbf{g}\in\mathcal{T}\left|\mathcal{Q}\mathbf{g}\in\mathcal{P}(H)\right.\right\} .\]

\end{defn}
\selectlanguage{english}%

%
{}
\begin{thm}
[\cite{Kim-isit10}]\foreignlanguage{american}{\label{thm:JointLP}\emph{The
LP joint decoder computes\begin{equation}
\argmin_{\mathbf{g}\in\mathcal{P}_{\mathcal{T}}(H)}\sum_{i\in\mathcal{I}}\sum_{\substack{e\in\mathcal{T}_{i}}
}b_{i,e}g_{i,e}\label{eq:JointLP}\end{equation}
and outputs a joint ML edge-path if $\mathbf{g}$ is integral.}}%
\begin{figure}
\selectlanguage{american}%
\framebox{\begin{minipage}[c]{0.97\columnwidth}%
\selectlanguage{english}%
\vspace{2mm}

\selectlanguage{american}%
\textbf{Problem-P:}\\
\begin{minipage}[t]{1\columnwidth}%
\vspace{-7mm}\[
\min_{\mathbf{g,w}}\sum_{i\in\mathcal{I}}\sum_{e\in\mathcal{T}_{i}}b_{i,e}g_{i,e}\]

\selectlanguage{english}%
\vspace{-5mm}

\selectlanguage{american}%
subject to\[
\sum_{\mathcal{B}\in\mathcal{E}_{j}}w_{j,\mathcal{B}}=1,\,\,\,\forall j\in\mathcal{J},\,\,\,\sum_{\substack{e\in\mathcal{T}_{p}}
}g_{p,e}=1,\,\mbox{for any}\, p\in\mathcal{I}\]
\[
\sum_{\mathcal{B}\in\mathcal{E}_{j},\mathcal{B}\ni i}w_{j,\mathcal{B}}=\sum_{e:x(e)=1}g_{i,e},\,\,\,\forall i\in\mathcal{I},j\in\mathcal{N}\left(i\right)\]

\[
\sum_{e:s'(e)=k}g_{i,e}=\sum_{e:s(e)=k}g_{i+1,e},\,\,\,\forall i\in\mathcal{I}\setminus N,\, k\in\emph{S}\]
\[
w_{j,\mathcal{B}}\geq0,\,\,\,\forall j\in\mathcal{J},\,\mathcal{B}\in\mathcal{E}_{j},\,\,\, g_{i,e}\geq0,\,\,\,\forall i\in\mathcal{I},\, e\in\mathcal{T}_{i}.\]

\vspace{-1mm}%
\end{minipage}%
\end{minipage}}\selectlanguage{english}

\end{figure}

\end{thm}
%
{}

\section{New Results: Iterative Joint LP Decoder \label{sec:Joint-LPD}}

\subsection{Iterative Joint LP Decoder Derivation}

\selectlanguage{american}%
%
{}In this section, we develop an iterative solver for the joint decoding
LP. There are few key steps in deriving our iterative solution for
the joint LP decoding problem. For the first step, given by Problem-P,
we reformulate the original LP \eqref{eq:JointLP} in Thm. \ref{thm:JointLP}
using only equality constraints involving the indicator variables%
\footnote{The valid patterns $\mathcal{E}_{j}\triangleq\left\{ \mathcal{B}\subseteq\mathcal{N}\left(j\right)|\left|\mathcal{B}\right|\,\mbox{is}\,\mbox{even}\right\} $
for each parity-check $j\in\mathcal{J}$ allow us to define the indicator
variables $w_{j,\mathcal{B}}$ (for $j\in\mathcal{J}$ and $\mathcal{B}\in\mathcal{E}_{j}$)
which equal 1 if the codeword satisfies parity-check $j$ using configuration
$\mathcal{B}\in\mathcal{E}_{j}.$%
} $\mathbf{g}$ and $\mathbf{w}$. 

\begin{figure}[t]
\begin{centering}
\includegraphics[scale=0.6]{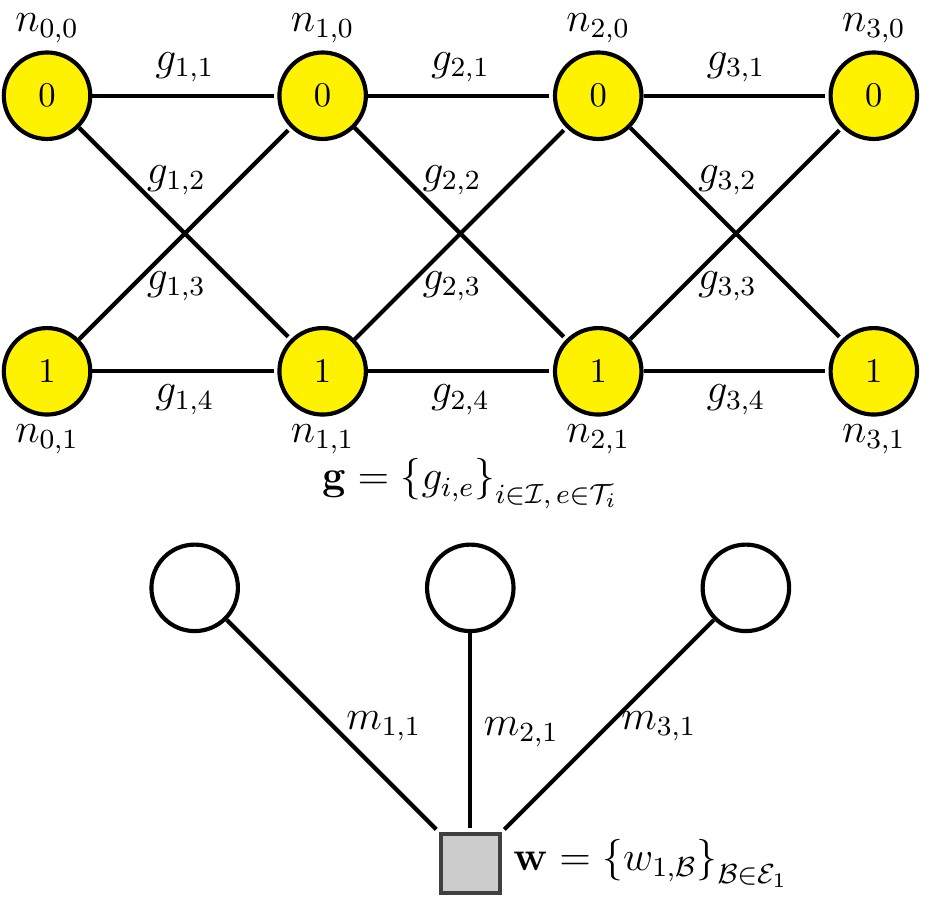}
\par\end{centering}

\caption{\label{fig:ProblemP}Illustration of primal variables $\mathbf{g}$
and $\mathbf{w}$ defined for Problem-P and dual variables $\mathbf{n}$
and $\mathbf{m}$ defined for Problem-D1 on the same example given
by Fig.  \ref{fig:Example}: SPC(3,2) with DIC for $N=3.$}

\end{figure}

The second step, given by Problem-D1, follows from standard convex
analysis (e.g., see \cite{Boyd-2004}). The Lagrangian dual of Problem-P
is equivalent to Problem-D1 and the minimum of Problem-P is equal
to the maximum of Problem-D1. From now on, we consider the Problem-D1
where the code and trellis constraints separate into two terms in
the objective function. See Fig. \ref{fig:ProblemP} for a diagram
of the variables involved.

The third step, given by Problem-D2, observes that forward/backward
recursions can be used to perform the optimization over $\mathbf{n}$
and remove one of the dual variable vectors. This splitting was enabled
by imposing the trellis flow normalization constraint in Problem-P
only at one time instant $p\in\mathcal{I}$. This detail gives $N$
different ways to write the same LP and is an important part of obtaining
update equations similar to TE. %
\begin{figure}
\framebox{\begin{minipage}[c]{0.97\columnwidth}%
\selectlanguage{english}%
\vspace{2mm}

\selectlanguage{american}%
\textbf{Problem-D1:}\\
\begin{minipage}[t]{1\columnwidth}%
\selectlanguage{english}%
\vspace{-5mm}\foreignlanguage{american}{\[
\max_{\mathbf{m,n}}\sum_{j\in\emph{J}}\min_{\mathcal{B}\in\mathcal{E}_{j}}\!\left[\sum_{i\in\mathcal{B}}m_{i,j}\right]\!\!+\!\min_{e\in\mathcal{T}_{p}}\!\left[\Gamma_{p,e}\!-\! n_{p-1,s(e)}\!+\! n_{p,s'(e)}\right]\]
subject to \vspace{-0mm}\[
\Gamma_{i,e}\geq n_{i-1,s(e)}-n_{i,s'(e)},\,\forall i\in\mathcal{I}\setminus p,\, e\in\mathcal{T}_{i}\vspace{-2mm}\]
and \vspace{-4mm}}

\selectlanguage{american}%
\[
n_{0,k}=n_{N,k}=0,\,\forall k\in\emph{S},\vspace{-2mm}\]

where \vspace{-4mm}

\emph{\[
\Gamma_{i,e}\triangleq b_{i,e}-\delta_{x(e)=1}\sum_{j\in\mathcal{N}(i)}m_{i,j}.\]
}%
\end{minipage}%
\end{minipage}}
\end{figure}

\selectlanguage{english}%

\begin{figure}[t]
\begin{centering}
\includegraphics[scale=0.65]{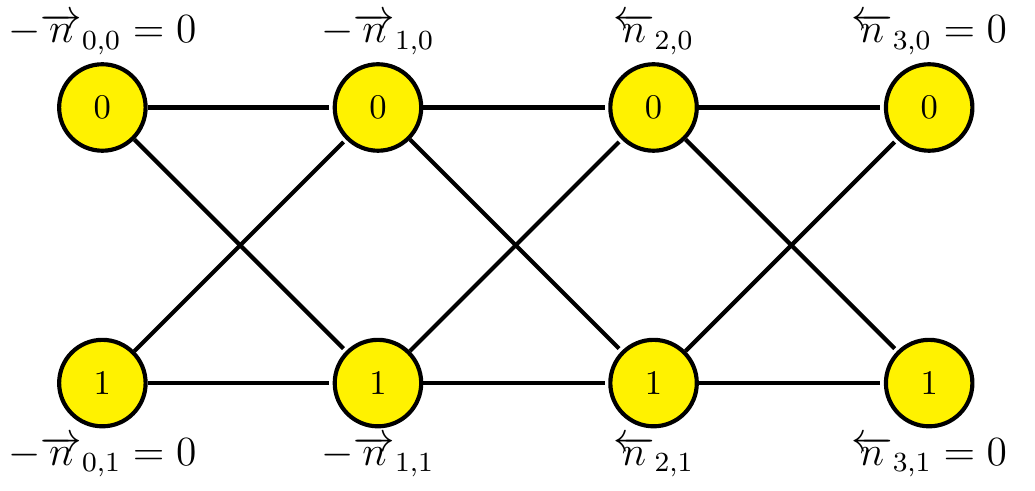}
\par\end{centering}

\caption{\label{fig:ProblemD-2}Illustration of Viterbi updates in Problem-D2
on the same example given by Fig. \ref{fig:Example}: DIC for $N=3$
with forward $\protect\overrightarrow{\mathbf{n}}$ and backward $\protect\overleftarrow{\mathbf{n}}.$ }

\end{figure}

\selectlanguage{american}%
\begin{lem}
\emph{\label{lem:Problem-D-2}Problem-D1 is equivalent to the Problem-D2.}\end{lem}
\begin{IEEEproof}
By rewriting the inequality constraint in Problem-D as\[
-n_{i,s'(e_{i})}\leq-n_{i-1,s(e_{i})}+\Gamma_{i,e}\]
we obtain the recursive upper bound for $i=p-1$ as\begin{align*}
 & -n_{p-1,k}\\
 & \leq\negthinspace\left.-n_{p-2,s(e_{p-1})}+\Gamma_{p-1,e}\right|_{s'(e_{p\!-\!1})=k}\\
 & \leq\negthinspace\left.-n_{p-3,s(e_{p-2})}\negthinspace+\negthinspace\Gamma_{p-2,e}\right|_{s'(e_{p\!-\!2})=s(e_{p\!-\!1})}\negthinspace+\negthinspace\left.\Gamma_{p-1,e}\right|_{s'(e_{p\!-\!1})=k}\\
 & \,\,\,\,\,\,\,\,\,\,\,\,\,\,\,\,\,\,\,\,\,\,\,\,\,\,\,\,\,\,\,\,\,\,\,\,\,\,\,\,\,\,\,\,\,\,\,\,\,\,\,\,\,\,\,\,\vdots\,\,\,\,\,\,\,\,\,\,\,\,\,\,\,\,\,\,\,\,\,\,\,\,\,\,\,\,\,\,\,\,\,\,\\
 & \leq\negthinspace\left.-n_{1,s(e_{2})}\negthinspace+\negthinspace\sum_{i=2}^{p-1}\Gamma_{i,e_{}}\right|_{s'(e_{p\!-\!1})=k,\negthinspace s'(e_{p\!-\!2})=s(e_{p\!-\!1}),\negthinspace\ldots,\negthinspace s'(e_{1})=s(e_{2}).}\end{align*}
This upper bound $-n_{p-1,k}\leq-\overrightarrow{n}_{p-1,k}$ is achieved
by the forward Viterbi update in Problem-D2 for $i=1,\,\ldots\,,\, p-1.$
Again, by expressing the same constraint as \[
n_{i-1,s(e_{i})}\leq\Gamma_{i,e}+n_{i,s'(e_{i})}\]
we get recursive upper bound for $i=p+1$. Similar reasoning shows
this upper bound $n_{p,k}\leq\overleftarrow{n}_{p,k}$ is achieved
by the backward Viterbi update in Problem-D2 for $i=N-1,N-2,\,\ldots\,,\, p.$
See Fig. \ref{fig:ProblemD-2} for a graphical depiction of this.
\end{IEEEproof}
The fourth step, given by Problem-DS, is replacing minimum operator
in Problem-D2 with the soft-minimum operation. A smooth approximation
is obtained by using \[
\mbox{min}\left(x_{1},\, x_{2},\,\ldots\,,\, x_{m}\right)\approx-\frac{1}{K}\ln\sum_{i=1}^{m}\texttt{e}^{-Kx_{i}}\]
 as in \cite{Vontobel-turbo06}. It is easy to verify that this log-sum-exp
function converges to the minimum function as $K$ increases. Since
the soft-minimum function is used in two different ways, we use different
constants, $K_{1}$ and $K_{2},$ for the code and trellis terms.
This Problem-DS allows one to to take derivative of \eqref{eq:JLP3-1}
(giving the KKT equations, derived in Lemma \ref{lem:InnerLoop}),
and represent \eqref{eq:JLP3-2} and \eqref{eq:JLP3-3} as BCJR-like
forward/backward recursions (given by Lemma \ref{lem:OuterLoop}).%
\begin{figure}
\framebox{\begin{minipage}[c]{0.97\columnwidth}%
\selectlanguage{english}%
\vspace{2mm}

\selectlanguage{american}%
\textbf{Problem-D2:}\\
\begin{minipage}[t]{1\columnwidth}%
\selectlanguage{english}%
\vspace{-5mm}\foreignlanguage{american}{\[
\max_{\mathbf{m}}\sum_{j\in\emph{J}}\min_{\mathcal{B}\in\mathcal{E}_{j}}\!\left[\sum_{i\in\mathcal{B}}m_{i,j}\right]\!+\min_{e\in\mathcal{T}_{p}}\!\left[\Gamma_{p,e}\!-\!\overrightarrow{n}_{p\!-\!1,s(e)}\!+\!\overleftarrow{n}_{p,s'(e)}\right]\]
where $\overrightarrow{n}_{i,k}$ is defined for $i=1,\,\ldots\,,\, p-1$
by \vspace{0mm} \[
-\overrightarrow{n}_{i,k}=\min_{e\in s'^{-1}(k)}-\overrightarrow{n}_{i-1,s(e_{i})}+\Gamma_{i,e},\,\forall k\in\mathcal{S}\vspace{-1mm}\]
and $\overleftarrow{n}_{i,k}$ is defined for $i=N-1,N-2,\,\ldots\,,\, p$
by \vspace{0mm}\emph{\[
\overleftarrow{n}_{i,k}=\min_{e\in s^{-1}(k)}\overleftarrow{n}_{i+1,s'(e_{i+1})}+\Gamma_{i+1,e},\,\forall k\in\mathcal{S}\vspace{-1.5mm}\]
}starting from \vspace{-1mm}\[
\overrightarrow{n}_{0,k}=\overleftarrow{n}_{N,k}=0,\,\forall k\in\mathcal{S}.\]
 \vspace{-5mm}}\selectlanguage{american}
\end{minipage}%
\end{minipage}}
\end{figure}

\begin{lem}
\emph{\label{lem:InnerLoop}The unique maximum of \eqref{eq:JLP3-1}
over $\left\{ m_{p,j'}\right\} _{j'\in\mathcal{N}(p)}$ can be found
using the KKT equations, and an iterative solution for} \emph{$p\in\mathcal{I}$
is given by }\textup{\[
m_{p,j'}=M_{p,j'}+\frac{\gamma_{p}}{K_{1}},\,\,\, M_{p,j'}\triangleq\frac{1}{K_{1}}\,\ln\,\frac{1-l_{p,j'}}{1+l_{p,j'}}\]
}\emph{for }$j'\in\mathcal{N}(p)$ \textup{where} \textup{\[
l_{p,j'}\triangleq\prod_{i\in\mathcal{N}(j')\setminus p}\tanh\left(\frac{K_{1}m_{i,j'}}{2}\right),\]
\[
\gamma_{p}\triangleq\ln\,\frac{\sum_{e\in\mathcal{T}_{p}:x(e)=0}\texttt{e}^{-K_{2}\left(\Gamma_{p}-\overrightarrow{n}_{p-1,s(e)}+\overleftarrow{n}_{p,s'(e)}\right)}}{\sum_{e\in\mathcal{T}_{p}:x(e)=1}\texttt{e}^{-K_{2}\left(\Gamma_{p}-\overrightarrow{n}_{p-1,s(e)}+\overleftarrow{n}_{p,s'(e)}\right)}}.\]
}%
{}
\end{lem}
\selectlanguage{english}%

\selectlanguage{american}%
\begin{figure}
\framebox{\begin{minipage}[t]{0.97\columnwidth}%
\selectlanguage{english}%
\vspace{2mm}

\selectlanguage{american}%
\textbf{Problem-DS:}\\
\begin{minipage}[t]{1\columnwidth}%
\selectlanguage{english}%
\noindent \vspace{-5mm}\foreignlanguage{american}{\begin{eqnarray}
\max_{\mathbf{m}}\,\,-\frac{1}{K_{1}}\sum_{j\in\emph{J}}\ln\sum_{\mathcal{B}\in\mathcal{E}_{j}}\texttt{e}^{-K_{1}\left\{ \sum_{i\in\mathcal{N}(j)}m_{i,j}\mathbbm{1}_{\mathcal{B}}(i)\right\} }\label{eq:JLP3-1}\\
-\frac{1}{K_{2}}\ln\sum_{e\in\mathcal{T}_{p}}\texttt{e}^{-K_{2}\left\{ \Gamma_{p,e}-\overrightarrow{n}_{p-1,s(e)}+\overleftarrow{n}_{p,s'(e)}\right\} }\nonumber \end{eqnarray}
where $\mathbbm{1}_{\mathcal{B}}\left(i\right)$ is the indicator
function of the set $\mathcal{B},$ $\overrightarrow{n}_{i,k}$ is
defined for $i=1,\,\ldots\,,\, p-1$ by\emph{ }\begin{equation}
-\overrightarrow{n}_{i,k}=-\frac{1}{K_{2}}\ln\!\!\sum_{e_{i}\in s'^{-1}(k)}\!\!\texttt{e}^{-K_{2}\left\{ -\overrightarrow{n}_{i-1,s(e_{i})}+\Gamma_{i,e}\right\} },\label{eq:JLP3-2}\end{equation}
and $\overleftarrow{n}_{i,k}$ is defined for $i=N-1,N-2,\,\ldots\,,\, p$
by\begin{align}
\overleftarrow{n}_{i,k} & =-\frac{1}{K_{2}}\ln\!\!\!\sum_{e_{i+1}\in s^{-1}(k)}\!\!\!\texttt{e}^{-K_{2}\left\{ \overleftarrow{n}_{i+1,s'(e_{i+1})}+\Gamma_{i+1,e}\right\} }\label{eq:JLP3-3}\end{align}
}

\noindent \vspace{-3mm}\foreignlanguage{american}{starting from\[
\overrightarrow{n}_{0,k}=\overleftarrow{n}_{N,k}=0,\,\forall k\in\mathcal{S}.\]
}

\vspace{-1mm}\selectlanguage{american}
\end{minipage}%
\end{minipage}}
\end{figure}

\begin{lem}
\label{lem:OuterLoop}\emph{ Equations \eqref{eq:JLP3-2} and \eqref{eq:JLP3-3}
are equivalent to the BCJR-based forward and backward recursion given
by (\ref{eq:BCJR1}), (\ref{eq:BCJR2}), and (\ref{eq:BCJR3}).}%
{}
\end{lem}
Now, we have all the pieces to complete the algorithm. As the last
step, we combine the results of Lemma \ref{lem:InnerLoop} and \ref{lem:OuterLoop}
to obtain the iterative solver for the joint decoding LP, which is
summarized in Algorithm \ref{alg:IJLP}.%
{}%
{}
\begin{rem}
\label{rem:TEconnection}While resulting Algorithm \ref{alg:IJLP}
has the bit-node update different from standard belief propagation
(BP), we note that setting $K_{1}=1$ in the inner loop gives the
exact BP check-node update and setting $K_{2}=1$ in the outer loop
gives the exact BCJR channel update. In fact, one surprising result
of this work is that such a small change to the BCJR-based TE update
provides an iterative solver for the LP whose complexity similar to
TE. It is also possible to prove the convergence of a slightly modified
iterative solver that is based on a less efficient update schedule.%
{}%
\begin{algorithm}
\caption{\label{alg:IJLP}Iterative Joint Linear-Programming Decoding}

\begin{itemize}
\item Step 1. Initialize $m_{i,j}=0$ for $i\in\mathcal{I},\, j\in\mathcal{N}\left(i\right)$
and iteration count $\ell=0.$
\item Step 2. Update Outer Loop: For $i\in\mathcal{I}$,

\begin{itemize}
\item (i) Compute bit-to-trellis message\vspace{-0.5mm} \[
\lambda_{i,e}=\texttt{e}^{-K_{_{2}}\Gamma_{i,e}}\vspace{-2mm}\]
where \[
\Gamma_{i,e}=b_{i,e}-\delta_{x(e)=1}\sum_{j\in\mathcal{N}(i)}m_{i,j}.\vspace{-1mm}\]

\item (ii) Compute forward/backward trellis messages\vspace{-0.5mm}

\begin{equation}
\!\!\!\!\!\!\!\!\!\!\!\!\!\!\!\!\!\!\alpha_{i+1}\left(k\right)\!=\!\frac{\sum_{e\in s'^{-1}(k)}\alpha_{i}\left(s(e)\right)\cdot\lambda_{i+1,e}}{\sum_{k}\sum_{e\in s'^{-1}(k)}\alpha_{i}\left(s(e)\right)\cdot\lambda_{i+1,e}}\label{eq:BCJR1}\end{equation}
\begin{equation}
\!\beta_{i-1}\left(k\right)\!=\!\frac{\sum_{e\in s^{-1}(k)}\beta_{i}\left(s'(e)\right)\cdot\lambda_{i,e}}{\sum_{k}\sum_{e\in s^{-1}(k)}\beta_{i}\left(s'(e)\right)\cdot\lambda_{i,e}},\label{eq:BCJR2}\end{equation}
where $\beta_{N}\left(k\right)=\alpha_{0}\left(k\right)=1/\left|\mathcal{S}\right|$
for all $k\in\mathcal{S}$. 

\item (iii) Compute trellis-to-bit message $\gamma_{i}$\vspace{-1mm}\begin{equation}
\!\!\!\!\!\!\!\!\gamma_{i}\!=\!\mbox{log}\,\frac{\sum_{e\in\mathcal{T}_{i}:x(e)=0}\alpha_{i-1}\left(s(e)\right)\lambda_{i,e}\beta_{i}\left(s'(e)\right)}{\sum_{e\in\mathcal{T}_{i}:x(e)=1}\alpha_{i-1}\left(s(e)\right)\lambda_{i,e}\beta_{i}\left(s'(e)\right)}\label{eq:BCJR3}\end{equation}

\end{itemize}
\item Step 3. Update Inner Loop for $\ell_{\mbox{inner}}$ rounds: For $i\in\mathcal{I}$,

\begin{itemize}
\item (i) Compute bit-to-check msg $m_{i,j}$ for $j\in\mathcal{N}\left(i\right)$\vspace{0.5mm}
\[
m_{i,j}=M_{i,j}+\frac{\gamma_{i}}{K_{1}}\]

\item (ii) Compute check-to-bit msg $M_{i,j}$ for $j\in\mathcal{N}\left(i\right)$\vspace{0.5mm}
\[
M_{i,j}=\frac{1}{K_{1}}\,\ln\,\frac{1-l_{i,j}}{1+l_{i,j}}\]
where\vspace{0mm} \[
l_{i,j}=\prod_{r\in\mathcal{N}\left(j\right)\setminus i}\tanh\left(\frac{K_{1}m_{r,j}}{2}\right)\]

\end{itemize}
\item Step 4. Compute hard decisions and stopping rule \vspace{0mm}

\begin{itemize}
\item (i) For $i\in\mathcal{I}$,\vspace{0mm} \begin{align*}
\hat{f}_{i} & =\begin{cases}
1 & \mbox{if}\,\,\,\gamma_{i}<0\\
0, & \mbox{otherwise}\end{cases}\end{align*}

\item (ii) If $\mathbf{\hat{f}}$ satisfies all parity checks or the iteration
number, $\ell_{\mbox{outer}}$, is reached, stop and output $\mathbf{\hat{f}}$.
Otherwise increase $\ell$ by 1 and go to Step 2.
\end{itemize}
\end{itemize}

\end{algorithm}

\end{rem}

\subsection{Convergence Analysis \label{sub:Convergence}}

This section considers the convergence properties of the proposed
Algorithm \ref{alg:IJLP}. Although we have always observed convergence
of Algorithm \ref{alg:IJLP} in simulation, our proof requires a modified
update schedule that is less computationally efficient. %
{} Following Vontobel's approach in \cite{Vontobel-turbo06}, which
is based on general properties of Gauss-Seidel-type algorithms for
convex minimization, we show that the modified version Algorithm \ref{alg:IJLP}
is guaranteed to converge. Moreover, a feasible primal solution can
be obtained that is arbitrarily close to the solution of Problem-P.
%
{}

The modified update rule for Algorithm \ref{alg:IJLP} consists of
cyclically, for each $p=1,\ldots,N$, computing the quantity $\gamma_{p}$
(via step 2 of Algorithm \ref{alg:IJLP}) and then updating $m_{p,j}$
for all $j\in\mathcal{N}(p)$ (based on step 3 of Algorithm \ref{alg:IJLP}).
The drawback of this approach is that one BCJR update is required
for each bit update, rather than for $N$ bit updates. This modification
allows us to interpret Algorithm \ref{alg:IJLP} as a Gauss-Seidel-type
algorithm. Therefore, the next theorem can be seen as a natural generalization
of \cite{Vontobel-turbo06}\cite{Burshtein-it09}.%
{}

\begin{figure}[t]
\framebox{\begin{minipage}[t]{0.97\columnwidth}%
\selectlanguage{english}%
\vspace{2mm}

\selectlanguage{american}%
\textbf{Problem-PS:}\\
\begin{minipage}[t]{1\columnwidth}%
\selectlanguage{english}%
\vspace{-5mm}\foreignlanguage{american}{\[
\min_{\mathbf{g,w}}\sum_{i\in\mathcal{I}}\sum_{e\in\mathcal{T}_{i}}b_{i,e}g_{i,e}-\frac{1}{K_{1}}\sum_{j\in\mathcal{J}}H(w_{j})-\frac{1}{K_{2}}H(g_{p})\]
}\selectlanguage{american}
\end{minipage}

subject to the same constraints as Problem-P. \vspace{1mm}%
\end{minipage}}
\end{figure}

\begin{thm}
\emph{\label{thm:Fesability}Let $P^{*}$ and $\tilde{P}$ be the
minimum value of Problem-P and Problem-PS}%
\footnote{To show connections between problem descriptions clearly, we write
the Lagrangian dual of Problem-DS as Problem-PS by letting $w_{j}\triangleq\left\{ w_{j,\mathcal{B}}\right\} _{\mathcal{B}\in\mathcal{E}_{j}}$,
$g_{p}\triangleq\left\{ g_{p,e}\right\} _{e\in\mathcal{T}_{p}}$ and
$H(x)\triangleq-\sum_{i}x_{i}\,\ln x_{i}$ for $x$ in the standard
simplex. The minimum of Problem-PS is equal to the maximum of Problem-DS.\emph{ }%
}\emph{ and denote $\tilde{\mathbf{g}}$ be the optimum solution of
Problem-PS. For any $\delta>0$},\emph{ there exist sufficiently large
$K_{1}$ and $K_{2}$ such that sufficient many iterations of the
modified Algorithm \ref{alg:IJLP} yields $\mathbf{\tilde{g}}$ which
is feasible in Problem-P and satisfies }\[
0\leq\frac{\tilde{P}-P^{*}}{N}\leq\delta.\]
%
{}
\end{thm}
\selectlanguage{english}%

\section{Performance of Iterative Joint LP Decoder\label{sec:Simulations} }

\selectlanguage{american}%
To validate proposed solutions for the problem of the joint decoding
of a binary-input FSC and outer LDPC, we use the following two simulation
setups: 
\begin{itemize}
\item For preliminary studies, we use $(3,\,5)$-regular binary LDPC codes
with length 455 on the precoded dicode channel (pDIC)
\item For practical study, we use a $(3,\,27)$-regular binary LDPC code
with length 4923 and rate 8/9 on the class-II Partial Response (PR2)
channel used as a partial-response target for perpendicular magnetic
recording. 
\end{itemize}
All parity-check matrices were chosen randomly except that double-edges
and four-cycles were avoided. Since the performance depends on the
transmitted codeword, the results were obtained for a few chosen codewords
of fixed weight. The weight was chosen to be roughly half the block
length, giving weights 226 and 2462. 

\begin{figure}[t]
\begin{centering}
\includegraphics[width=0.3\paperheight]{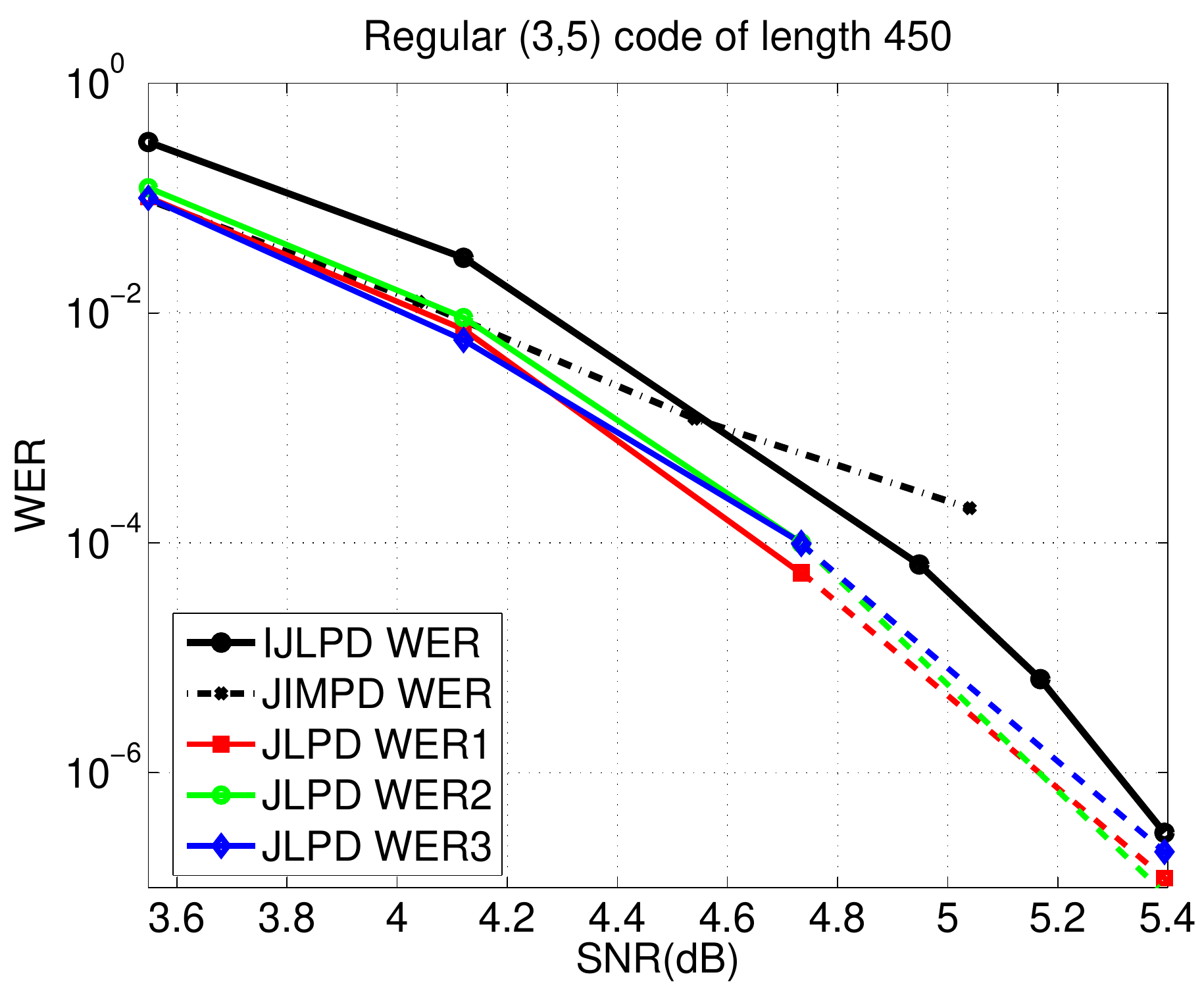}\foreignlanguage{english}{\vspace{0mm}}
\par\end{centering}

\selectlanguage{english}%
\caption{\label{fig:result}\foreignlanguage{american}{\label{fig:result1}This
figure shows comparison between the joint LP decoding (JLPD), joint
iterative message-passing decoding (JIMPD), and iterative joint LP
decoding (IJLPD) on the pDIC with AWGN for random (3,5) regular LDPC
codes of length} $N=450$. \foreignlanguage{american}{The curves shown
are the JLPD WER (solid), JLPD WER prediction (dashed), JIMPD WER
(dash-dot), and IJLPD WER (circle-solid).} The dashed curves are computed
using the union bound based on joint-decoding pseudo-codewords observed
at 2.67 dB as described in \cite{Kim-isit10}\foreignlanguage{american}{
and the dash-dot curves are obtained using the state-based JIMPD described
in \cite{Kavcic-it03}. The circle-solid curves are computed using
Algorithm \ref{alg:IJLP}.} Note that SNR is defined as channel output
power divided by $\sigma^{2}$.}
\vspace{0mm}\selectlanguage{american}

\end{figure}

Fig. \ref{fig:result1} shows the decoding results based on the Algorithm
\ref{alg:IJLP} compared with the joint LP decoding performed in the
dual domain using MATLAB in the first setup. The choice of parameters
and scheduling scheme has yet to be optimized. Instead, we use a simple
scheduling update scheme: variables are updated cyclically with 5
inner loop iterations after single outer iteration with $K_{1}=1000\,\mbox{and}\, K_{2}=100$.
Somewhat interestingly, we find that iterative joint LP decoding WER
curve loses about 0.2~dB at low SNR. This may be caused by using
too few iterations or finite values of $K_{1}$ and $K_{2}$. But,
at high SNR this gap disappears and the curve converges towards the
error rate predicted for joint LP decoding. This shows that joint
LP decoding outperforms belief-propagation decoding for short length
code at moderate SNR with the predictability of LP decoding. Of course
this can be achieved with a computational complexity similar to turbo
equalization. 

\begin{figure}[t]
\selectlanguage{american}%
\begin{centering}
\includegraphics[width=0.3\paperheight]{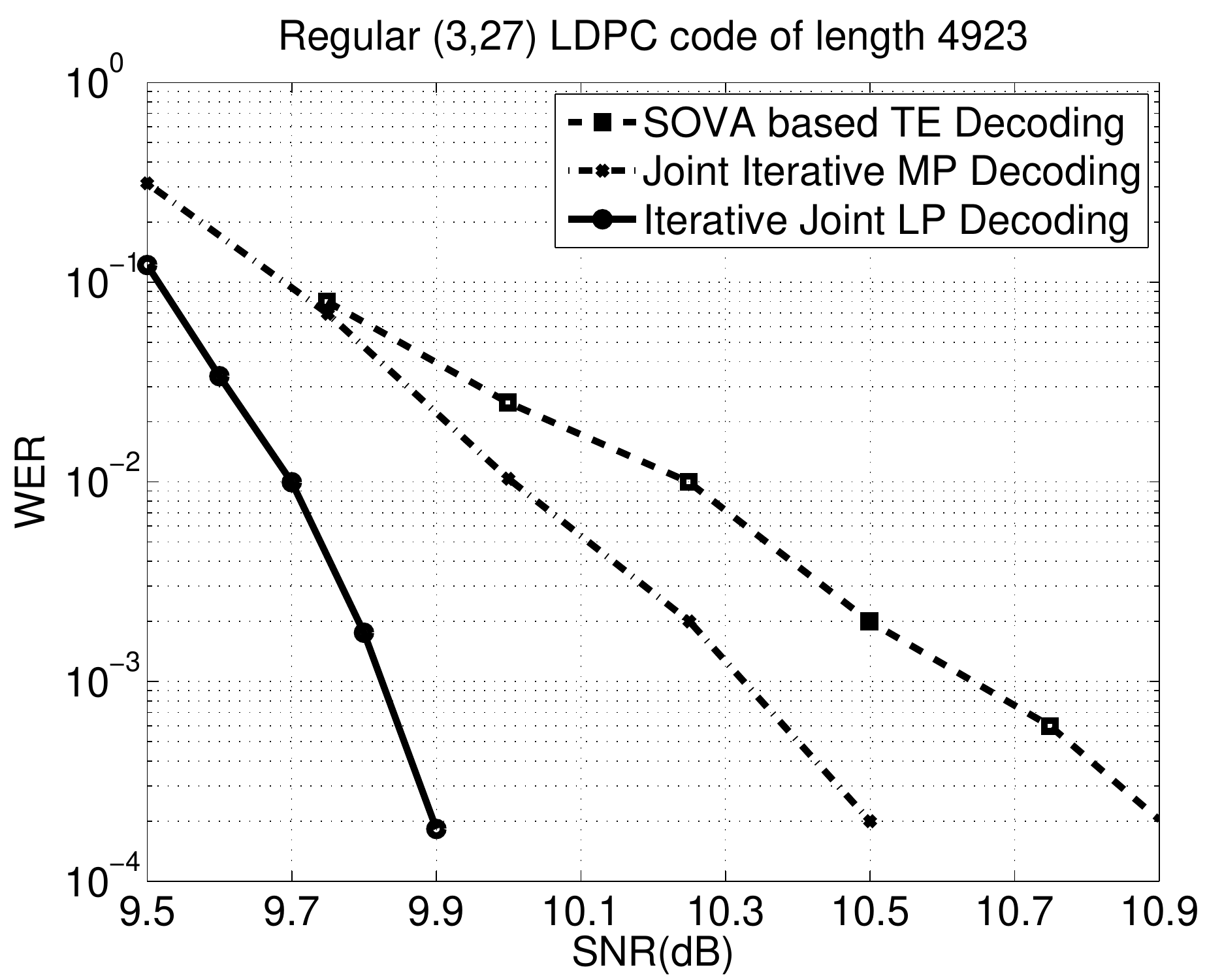}
\par\end{centering}

\caption{\label{fig:result3}This figure shows comparison between the iterative
joint LP decoding and other TE based methods on the PR2 channel with
AWGN for random (3,27) regular LDPC codes of length $N=4923$. The
curves shown are the joint iterative LP decoding WER (solid), the
stated-based joint iterative message-passing (MP) decoding WER (dash-dot)
described in \cite{Kavcic-it03}, and the soft output Viterbi algorithm
(SOVA)-based TE decoding WER (dashed) taken from \cite{Jeon-itm07}.
Note that SNR is defined as channel output power divided by $\sigma^{2}$.}
\selectlanguage{english}

\end{figure}

Fig. \ref{fig:result3} shows the decoding results based on the Algorithm
\ref{alg:IJLP} compared with the state-based JIMPD algorithm described
in \cite{Kavcic-it03} in more practical scenario. To make a fair
comparison, we fix the maximum iteration count, $\ell_{\mbox{outer}}\left(\ell_{\mbox{inner}}+1\right)$
of each algorithm to roughly 1000 and choose $K_{1}=1000\,\mbox{and}\, K_{2}=10$
for Algorithm \ref{alg:IJLP}. Surprisingly, we find that iterative
joint LP decoding WER curve with Algorithm \ref{alg:IJLP} wins over
JIMPD at all SNR with substantial gains. Also, the slope difference
between two curves anticipate greatly better error-floor performance
of Algorithm \ref{alg:IJLP}. This shows that joint LP decoding outperforms
belief-propagation decoding even for long length code at all SNR with
a computational complexity similar to TE.\foreignlanguage{english}{}%

\selectlanguage{english}%
\vspace{0mm}

\section{Conclusions \label{sec:Concl}}

\vspace{0mm}In this paper, we consider the problem of low-complexity
joint linear-programming (LP) decoding of low-density parity-check
codes and finite-state channels. We present a novel iterative solver
for the joint LP decoding problem. This greatly reduces the computational
complexity of the joint LP solver by exploiting the LP dual problem
structure. %
{}Its main advantage is that it provides the predictability of LP decoding
\foreignlanguage{american}{and significant gains over turbo equalization
(TE) with a computational complexity similar to TE.} \vspace{0mm}

\bibliographystyle{IEEEtran}

\foreignlanguage{american}{}%

\end{document}